\documentclass[aps,twocolumn]{revtex4}
\usepackage{epsfig,graphics,amssymb,amsmath,subeqnarray,txfonts}

\begin{document}

\title{Geometric transition in friction  for flow over a bubble mattress}
\author{Anthony M. J. Davis\footnote{Email: amdavis@ucsd.edu}}
\author{Eric Lauga\footnote{Email: elauga@ucsd.edu}}
\affiliation{Department of Mechanical and Aerospace Engineering, University of California San Diego, 9500 Gilman Drive, La Jolla CA 92093-0411, USA.}

\date{\today}
\begin{abstract}

Laminar flow over a bubble mattress is expected to experience a significant reduction in friction since the individual surfaces of the bubbles are shear-free. However, if the bubbles are sufficiently curved, their protrusion into the fluid and along the flow direction can lead to an increase in friction as was recently demonstrated experimentally and computationally. We provide in this paper a simple model for this result.  We consider a  shear flow at low Reynolds number past a two-dimensional array of bubbles, and calculate analytically the effective slip length of the surface as function of the bubble geometry in the dilute limit. Our model is able to  reproduce quantitatively the relationship between effective friction and bubble geometry obtained in numerical computations, and in particular: (a) The asymmetry in friction between convex and concave bubbles, and (b) the existence of a geometric transition from reduced to enhanced friction at a critical bubble protrusion angle. 

\end{abstract}
\maketitle

Microfluidic devices are used to manipulate small volumes of liquids (micro- and nano-liters), and have important applications in  biology, chemistry and engineering  \cite{stone_review,rmp}.  However, from a practical standpoint, one important caveat of decreasing length scales is the increase in friction.  For  flow driven by a pressure gradient, $\nabla P$, in a device of typical cross-sectional scale $a$, the volume flow rate scales as $Q\sim a^4 \nabla P/\eta$, where $\eta$ is the viscosity of the fluid. Consequently, imposing a constant flow rate in a device decreasing in size requires a sharp increase of the applied pressure gradient as $\nabla P\sim 1/a^4$. For example, driving a flow rate of one microliter of water per second in a $10$~$\mu$m-wide and $1$ cm-long  device requires a pressure drop  of many atmospheres. 

Viscous friction is therefore an important issue on small scales, and engineers have looked for ways to reduce it. Recently a number of groups have reported slip, or apparent slip, for simple  shear or pressure-driven flows over hydrophobic surfaces (see Refs.~\cite{neto05,laugareview,bocquet07} for a review). Slip is typically characterized by a slip length, $\lambda$, which is the fictional distance below the surface at which the velocity field would extrapolate to zero. 
For a shear flow with shear rate $\dot\gamma$ over a flat slipping surface,  the velocity field is written as
\begin{equation}\label{shearflow}
{\bf u} = (\lambda \dot\gamma + \dot\gamma y)\hat{\bf x}
\end{equation}
where $\hat{\bf x}$ is the flow direction, and $y$ the direction perpendicular to the surface.
For a system of typical cross-sectional size $a$,  friction decreases  as some increasing function of $\lambda/a$ \cite{neto05,laugareview,bocquet07,sbragaglia07,ybert07,steinberger07,hyvaluoma08}. Therefore for small enough systems, or large enough slip lengths, the friction reduction could be substantial.
However, intrinsic slip lengths of hydrophobic surfaces do not exceed tens  of nanometers \cite{neto05,laugareview,bocquet07}. As a result, the effect is small for devices on the micron-scale, and other means of friction-reduction must be devised. 

\begin{figure}[b]
\centering
\includegraphics[width=0.48\textwidth]{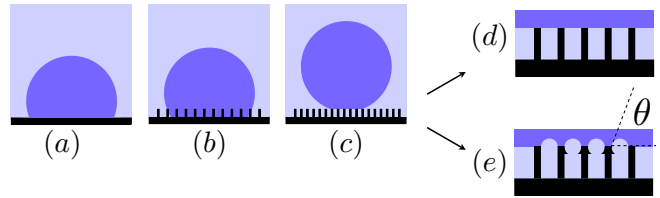}
\caption{Surperhydrophobic surfaces and bubble mattress: 
Liquid drop on a: (a)  flat hydrophobic surface; 
(b) rough hydrophobic surface in  Wenzel state; 
(c) rough hydrophobic surface  in Cassie state. 
When the drops sits partially on air (Cassie state), the gas-liquid interface can be flat (d)  or curved (e), with a protrusion angle $\theta$ measured with respect to the horizontal direction.}
\label{superhydrophobic}
\end{figure}

One idea recently discussed is to exploit super-hydrophobic surfaces  \cite{quere05,degennes_book} to reduce drag.  When an hydrophobic solid surface (Fig.~\ref{superhydrophobic}a) is sufficiently rough, under certain geometric   conditions, a low-pressure high-surface energy fluid in contact with the solid can spontaneously de-wet \cite{degennes_book}, thus transitioning from a Wenzel state where the fluid fills the grooves on the surface (Fig.~\ref{superhydrophobic}b), to a fakir-like Cassie state  where the fluid sits partially on the solid surface and partially on gas (Fig.~\ref{superhydrophobic}c) \cite{quere05,degennes_book}. This phenomenon can also occur at the nanometer scale \cite{cottin-bizonne03}. When de-wetting does occur, shear stresses on the portion of the fluid interface in contact with the gas are expected to be insignificant,  and the effective friction of such a super-hydrophobic surface should be significantly lower than that of a solid surface. Drag reduction by super-hydrophobic surfaces was indeed demonstrated experimentally, both for macro-scale pipe flow  \cite{watanabe99}, as well as microchannels \cite{ou04,ou05}.  Numerical simulations also show a similar effect at the molecular scale  \cite{cottin-bizonne03}.

Liquids in contact with a mixed solid/gas interface are therefore expected to display low friction. However, one issue can potentially limit the drag-reducing properties of such surfaces, namely the geometry of the gas-liquid interface. When the pressure in the gas is similar to that in the liquid above it, the interface  remains flat (Fig.~\ref{superhydrophobic}d), and drag is expected to be reduced by the gas-liquid interface. However, when the pressure in the gas is larger than that in the liquid, the interface becomes curved (Fig.~\ref{superhydrophobic}e), with mean curvature given by the Young-Laplace equation \cite{degennes_book}. In that case, the protrusion of the interface into the liquid  curtails the drag-reducing properties of the surface by distorting the flow streamlines in the shear direction  \cite{richardson73,jansons88,legendre08_1,legendre08_2}. Notably, if we denote by $\theta$ the value of the protrusion angle of the spherical-cap interface with respect to the horizontal surface (see Fig.~\ref{superhydrophobic}e), two recent experimental and numerical studies demonstrated that if the interface  extends into the liquid beyond a critical value $\theta_c$, the friction of the surface becomes larger than that of the original (flat) no-slip surface  \cite{steinberger07,hyvaluoma08}. These results imply therefore that the geometry of free surfaces plays a crucial role vis-a-vis the resistance to viscous flow, implying in particular that  the recently-discovered nano-bubbles  \cite{ishida00,tyrrell01} might not always lead to as low a friction as previously thought.

Here, we specifically address this geometric transition from reduced to enhanced friction for shear flow over a collection of bubbles. We consider a simple two-dimensional model to predict the critical value of the apparent angle $\theta_c$ of the bubbles on the (otherwise) horizontal surface at which the effective surface starts displaying a friction larger than that obtained by a smooth solid surface.  In the  two-dimensional case, the velocity field can be solved exactly in the dilute limit, and the geometric transition is seen to happen at $\theta_c\approx 65^\circ$. More generally the relationship between effective slip length and apparent angle from numerical simulations is well represented by our simple model~\cite{steinberger07,hyvaluoma08}.

\begin{figure}[t]
\centering
\includegraphics[width=0.4\textwidth]{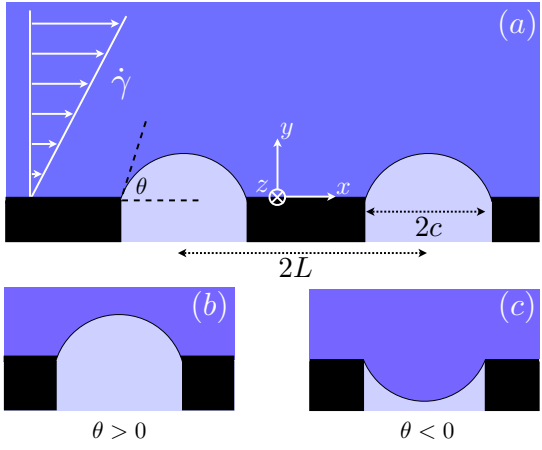}
\caption{Setup for the two-dimensional calculation and notation: (a) shear flow (shear rate, $\dot\gamma$) over a collection of bubbles of projected radius $c$, protrusion angle $\theta$ and separation distance $2L$. The calculation is performed asymptotically in the dilute limit, $\phi=c/L\ll 1$; (b): The case $\theta>0$ corresponds to a concave bubble protruding into the liquid; (c):  $\theta<0$ models a convex bubble curved away from the otherwise flat surface.}
\label{2D}
\end{figure}

We consider therefore the setup displayed in Fig.~\ref{2D}: A dilute collection of two-dimensional bubbles (no-shear surfaces) on an otherwise no-slip surface. We assume the capillary number to be sufficiently small that the bubbles have perfectly circular shape.
Their projected radius on the surface is $c$ and they are separated by a distance $2L$. The bubble surface coverage is therefore given by $\phi=c/L$. The protrusion angle of the the bubble into the fluid is denoted $\theta$; $\theta > 0$ is a concave bubble that protrudes into the fluid (Fig.~\ref{2D}b), whereas $\theta <0$ is a convex bubble protruding underneath the surface (Fig.~\ref{2D}c). The velocity field away from the surface is a pure shear flow, with constant shear rate $\dot\gamma$. We denote by $x$ the direction along the surface,  $y$ the direction of shear, and $z$ the third direction, and solve for the flow field in the limit of low Reynolds number  \cite{happel}.

We first examine the case of a single bubble. The velocity field, ${\bf v}$, is solved by introducing a streamfunction, ${\bf v}=\hbox{curl}(\psi\hat{\bf{z}})$, where $\nabla^4\psi=0$
in the half-plane $y>0$, indented by an arc of the circle 
$x^2+(y+c\cot\theta)^2=c^2csc^2\theta$ which perturbs the shear flow
$\psi=\frac{1}{2}\dot{\gamma}y^2$. There is perfect slip without penetration at 
the surface of the two-dimensional bubble, and no slip is enforced at the rigid plane $y=0$.  We introduce toroidal coordinates defined by
\begin{equation}\label{torco}
x=\frac{c\sinh\xi}{\cosh\xi+\cos\eta}, \quad  y=\frac{c\sin\eta}
{\cosh\xi+\cos\eta},  
\end{equation}
which is the conformal mapping $x+iy=c\tanh(\xi+i\eta)/2$. The fluid
region is then $-\infty<\xi<\infty,\theta<\eta<\pi$, with 
$\xi=\pm\infty$ at the special points $(\pm c,0)$, $\xi=0$ on the 
$y$-axis, $\eta=\theta$ on the arc and $\eta=\pi$ at $y=0, |x|>c$. A
suitable biharmonic stream function, asymptotic at infinity to the
prescribed shear flow, is given by \cite{davis77,Morse_book}
\begin{eqnarray}\label{psi}
\psi&=&\dot{\gamma}c^2\left[\frac{\frac{1}{2}\sin^2\eta}{(\cosh\xi+\cos\eta)^2}-\frac{\int_0^\infty f(s,\eta)\cos s\xi\, {\rm d} s}
{\cosh\xi+\cos\eta}\right],   \\
f&=&-A(s)\sin\eta\frac{\sinh s(\pi-\eta)}{s} \nonumber \\
&& +B(s)\left[\cos\eta
\frac{\sinh s(\pi-\eta)}{s}+\sin\eta\cosh s(\pi-\eta)\right].  \nonumber
\end{eqnarray}
Here, two functions have been eliminated by imposing no slip 
($\psi=0=\partial\psi/\partial\eta$ at $\eta=\pi$) on the rigid
plane on either side of the bubble. Note that using the method in Eq.~\eqref{psi} allows us to capture the appropriate stress singularity at the contact point between the bubble and the surface.

The first equation for the functions $A$ and $B$ is established by demanding that the arc
$\eta=\theta$ be a streamline on which $\psi=0$. Thus the transform,
\begin{equation}\label{Fourtran}
\int_0^\infty \frac{\sinh s\eta}{\sinh s\pi}\cos s\xi\, {\rm d} s=
\frac{\frac{1}{2}\sin\eta}{(\cosh\xi+\cos\eta)}, \quad (|\eta|<\pi),
\end{equation}
yields, by substituting Eq.~(\ref{psi}),
\begin{equation}\label{ABeqn1}
f(s,\theta)=\sin\theta\frac{\sinh s\theta}{\sinh s\pi}.
\end{equation}
The second equation  is established by demanding that the
strain rate component, $e_{\xi\eta}$, vanishes at $\eta=\theta$ in order to
achieve perfect slip. Since
\begin{equation}
2e_{\xi\eta}=\frac{\partial}{\partial\xi}\left(-\frac{1}{h^2}
\frac{\partial\psi}{\partial\xi}\right)+\frac{\partial}{\partial\eta}
\left(\frac{1}{h^2}\frac{\partial\psi}{\partial\eta}\right), \quad h=\frac{c}{\cosh\xi+\cos\eta},
\end{equation}
it is found that
\begin{eqnarray}
&& \int_0^\infty \left[\frac{\partial^2f}{\partial\eta^2}+(s^2+1)f \right]_{\eta=\theta}\cos s\xi \, {\rm d} s\\ &&=
\cos\theta\left[\frac{\cos\theta}
{\cosh\xi+\cos\eta}+\frac{4\sin^2\theta}{(\cosh\xi+\cos\eta)^2}\right]\\
&&+\sin\theta\left[-\frac{\sin\theta}{\cosh\xi+\cos\eta}+
\frac{2\sin^3\theta}{(\cosh\xi+\cos\eta)^3}\right].
\end{eqnarray}
The first and second derivatives of Eq.~(\ref{Fourtran}) give
\begin{equation}\label{ABeqn2}
\left[\frac{\partial^2f}{\partial\eta^2}+(s^2+1)f\right]_{\eta=\theta}
=\frac{2s[s\sin\theta\sinh s\theta+
\cos\theta\cosh s\theta]}{\sinh s\pi}.
\end{equation}
After substitution of Eq.~(\ref{psi}) into Eq.~(\ref{ABeqn1}) and 
Eq.~(\ref{ABeqn2}), it is found that
\begin{eqnarray}\label{eqA}
A(s)&=&\frac{s}{\sinh 2s(\pi-\theta)+s\sin 2\theta} \\
&& \times\left[\cos 2\theta+
\frac{s\sin 2\theta\cosh s\pi+\sinh s(\pi-2\theta)}{\sinh s\pi}\right],\nonumber
\end{eqnarray}
and
\begin{equation}\label{AB}
B(s)=\frac{s\sin 2\theta}{\sinh 2s(\pi-\theta)+s\sin 2\theta}.
\end{equation}

We can now derive the result for a period array of bubbles.
Away from the surface, the far-field velocity field is given, according to Eq.~(\ref{torco}), by 
$\xi^2+(\pi-\eta)^2\to 0$ ($\eta<\pi$), with 
\begin{equation}
x\sim\frac{2c\xi}{\xi^2+(\pi-\eta)^2}, \quad 
y\sim\frac{2c(\pi-\eta)}{\xi^2+(\pi-\eta)^2}. 
\end{equation}
Hence  the asymptotic form of the perturbation term in Eq.~(\ref{psi}) is given by
\begin{equation}\label{farfield}
2\dot{\gamma}c^2\frac{y^2}{x^2+y^2}\int_0^\infty A(s)ds. 
\end{equation}
In the presence of a periodic array of bubbles, with period $2L$,
the prefactor in Eq.~\eqref{farfield} is replaced by
$
\displaystyle \sum_{n=-\infty}^\infty\frac{y^2}{(x-2nL)^2+y^2}
$ with a mean   value  available from the identity
\begin{eqnarray}
&&\frac{1}{2L}\int_{-L}^L\sum_{n=-\infty}^\infty\frac{y^2dx}
{(x-2nL)^2+y^2}=\frac{y}{2L}\int_{-\infty}^\infty\frac{ydx}{x^2+y^2}
= \frac{\pi y}{2L}\cdot\quad\quad
\end{eqnarray}
Consequently,  the far field velocity  perturbation is given by
\begin{equation}
\pi \dot{\gamma}c\left(\frac{c}{L}\right)\int_0^\infty A(s)ds \,\hat{\bf x},
\end{equation}
which, with $A(s)$ given by Eq.~(\ref{eqA}), is identified as an effective slip speed at $y=0$ with dimensionless slip length (see Eq.~\ref{shearflow}).
\begin{equation}\label{final}
\frac{\lambda}{c}=\pi \left(\frac{c}{L}\right)\int_0^\infty A(s)ds.
\end{equation}

\begin{figure}[t]
\centering
\includegraphics[width=0.48\textwidth]{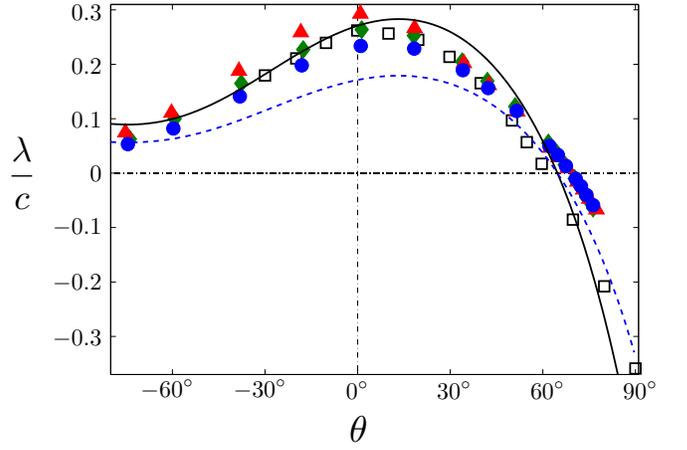}
\caption{(color online) Comparison between the numerical data of Ref.~\cite{steinberger07} (empty symbols) and \cite{hyvaluoma08} (filled symbols) and our two-dimensional dilute model: Dimensionless effective slip length as a function of the protrusion angle of the bubble into the fluid.
Symbols:  Square lattice ($\square$, $\phi=0.68$ \cite{steinberger07}; $\bullet$, $\phi=0.43$ \cite{hyvaluoma08}); rectangular lattice ($\blacklozenge$, $\phi=0.43$ \cite{hyvaluoma08}); rhombic lattice ($\blacktriangle$, $\phi=0.43$ \cite{hyvaluoma08}). Lines: Our two-dimensional dilute model, Eq.~\eqref{final}, for two different surface coverage,  $\phi=0.43$~(- - -) and $\phi=0.68$~(---).}
\label{2Dresults}
\end{figure}

Let us  now compare the predictions of our model (Eq.~\ref{final}) with the numerical computations of Refs~\cite{steinberger07,hyvaluoma08}. The results are displayed in Fig.~\ref{2Dresults}, where we plot the effective slip length of the surface (non-dimensionalized by the projected bubble radius on the surface, $c$) as a function of the protrusion angle, $\theta$, into the fluid. The simulations of Ref.~\cite{steinberger07} consider a three-dimensional square lattice, with a surface coverage of the bubbles  $\phi=0.68$ (squares). The computations of Ref.~\cite{hyvaluoma08} are for three different lattices (square, circles; rectangular lattice, lozenges; rhombic lattice, triangles) and in all cases where the dependence on the protrusion angle is studied the  fraction of the surface covered by the bubbles is $\phi=0.43$.  We also plot in Fig.~\ref{2Dresults} the results of our model for both surface coverage considered in the two previous studies: $\phi=0.43$ (dashed line) and $\phi=0.68$ (solid line) \footnote{Strictly speaking, our two-dimensional model was derived in the low-$\phi$ limit. We  consider however its results beyond its regime of asymptotic validity, as it is expected to provide a good quantitative model even in the limit of large $\phi$ and reproduce the essentially physical features of the phenomenon.}.

The main features of the full numerical results are seen to be reproduced by our analytical model. There exists a critical protrusion angle, $\theta_c$, above which the effect of the wall-attached bubbles displays a transition from reduced ($\theta<\theta_c$) to enhanced friction ($\theta > \theta_c$). Our model predicts $\theta_c\approx 65^\circ$, in good agreement with the results of Ref.~\cite{steinberger07} ($\theta_c\approx 62^\circ$) and Ref.~\cite{hyvaluoma08} ($\theta_c\approx 69^\circ$). Our model also successfully reproduces the asymmetry in the friction curves between $\theta<0$ and $\theta>0$, indicating a qualitative difference between the effect of convex and concave bubbles on shear flow. Experimentally, this asymmetry implies that flow over an array of bubbles  at a lower pressure than the fluid ($\theta<0$) is less likely to show an increase of wall friction than flow over an array of bubbles with a larger pressure than the fluid ($\theta>0$).

The model we consider here is two-dimensional. A simple argument can also be made to show that the critical protrusion angle lies in the range $0^\circ < \theta_c < 90^\circ$ in three dimensions.
For $\theta=0^\circ$, the bubbles are flat disks, for which the calculation is available in Ref.~\cite{davis91}, where it is found \footnote{After noting that, in his Eq.~(22) of Ref.~\cite{davis91}, the toroidal coordinate $\lambda\sim 
r$ in the far field and $\zeta=y/\lambda$.} that the 
asymptotic form of the perturbation flow component $v_x$ is
$2\dot{\gamma}c^3yx^2/3\pi r^{5}$, with $r^2=x^2+y^2+z^2$. In the presence of a (dilute) square lattice of such disks, with period $2L$, the effective  slip length is simply found to be
${\lambda}/{c}= {c^2}/9{L^2}$.
This is the same result as that obtained by Ref.~\cite{sbragaglia07}, and indicates that bubbles with $\theta=0^\circ$ reduce the effective friction of the surface, and therefore $0^\circ < \theta_c $.
In the case where $\theta=90^\circ$ (hemispherical bubble), the velocity field is readily verified to be given by
${\bf v}=\dot{\gamma}y\left[\hat{\bf{x}}-{c^3x\hat{\bf{r}}}/{r^4}\right]$ \cite{legendre08_1,legendre08_2}. 
In the presence of a square lattice of such bubbles, the perturbation velocity in the far field can be simply added up (in the dilute limit), and the obtained slip length is given by
${\lambda}/{c}=-{\pi c^2}/{6 L^2}$. The negative slip length indicates that the array of stress-free hemispheres increases the effective friction of these surfaces, and therefore $\theta_c < 90^\circ$.

In summary,  this paper presents a two-dimensional model of shear flow past an array of bubbles. We calculate the effective slip length of the surface in the dilute limit, and found good agreement with recent numerical three-dimensional numerical simulations for flow over super-hydrophobic surfaces. In particular, our results reproduce the asymmetry between convex and concave bubbles, as well as the existence of a geometric transition from low to high friction at a critical bubble protrusion angle. Our approach provides therefore the minimal model necessary to quantitatively capture the interplay 
between  low friction and geometry in shear flows. 

Our results could be extended by relaxing the various assumptions made in the paper, in particular by going beyond the  two-dimensional, dilute and zero Capillary number limits. In the case of three-dimensional bubbles, the issue of geometry of the lattice needs to be further explored as well.  Experimentally, the fabrication of controlled porous surfaces  \cite{zheng08} or pre-designed bubble lattices \cite{bremond06} would present exciting opportunities to map out and optimize the frictional properties of super-hydrophobic surfaces. In addition,  simulations showing an increase in friction  of a bubble mattress with capillary numbers  \cite{hyvaluoma08} could be addressed with a similar framework.

This work was supported in part by the National Science Foundation (Grants No. CTS-0624830 and CBET-0746285).

\bibliographystyle{unsrt}
\bibliography{bumps}
\end{document}